\documentstyle[12pt,version2,aps]{revtex}

\begin{document}

  \begin{title}
Quantum Group and Magnetic Translations.\\
 Bethe-Ansatz solution for  Bloch electrons in a magnetic field
\end{title}
   \author{P.B Wiegmann* and A.V.Zabrodin**}

\begin{instit}

* James Frank Institute and Enrico Fermi Institute of the University
of
Chicago,\\
5640 S.Ellis Ave.,Chicago Il
60637,e-mail:WIEGMANN@CONTROL.UCHICAGO.EDU
\\and
\\Landau Institute for Theoretical Physics\\
** Enrico Fermi Institute and Mathematical Disciplines Center of the
University
of Chicago,\\
5640 S.Ellis Ave.,Chicago Il 60637\\
and \\the Institute of Chemical Physics, Kosygina St. 4, SU-117334,
Moscow,
Russia \end{instit}

\begin{abstract}

We present a new approach to the problem of Bloch electrons in  magnetic
field,\\ by making explicit a natural relation between magnetic  translations
and the\\quantum group $U_{q}(sl_2)$.  The approach allows to express
the spectrum and\\\ the Bloch function as solutions of the Bethe-Ansatz
equations typical for com\\pletely integrable quantum systems
\\
\\
\\

\end{abstract}

\pacs{PACS numbers: 71,75, 02}


\section{Introduction}

 Several times a peculiar problem of Bloch electrons in
magnetic field emerged with a new face to describe another physical
application \cite{Z},\cite{A},\cite{W},\cite{H},\cite{TKNN}

i) it resembles some properties of the integer Hall effect
\cite{TKNN},

ii) its spectrum has an extremely rich structure of Cantor set, and
exibits
a {\it multifractal} behaviour \cite{H},\cite{AA}, \cite{T} ( see
also
\cite{K} for a review)

iii) it describes the localization phenomenon in {\it  quasiperiodic}
 potential (see e.g.\cite{K} and references therein).

iv) it has been recently conjectured that the symmetry of magnetic
group
may appear dynamically in strongly correlated electronic
systems\cite{Anderson},\cite{wiegmann}.

In this paper we show that this problem (some times called Hofstadter problem)
and a class of quasiperiodic
equations are solvable by the {\it Bethe- Ansatz}. We made explicit a long time
anticipated connection of the group of magnetic translations with the {\it
Quantum Group} $U_{q}(sl_2)$ and  with Quantum Integrable Systems. The result
of the paper is the algebraic {\it Bethe- Ansatz equations} for
the spectrum. Although we do not solve the Bethe-Ansatz equation
here, we are confident that they provide a basis for analytical study of
the multyfractal properties of the spectrum.
\section{Model and Result  }

The Hamiltonian of a particle on a two dimensional square lattice in
magnetic field is
\begin{equation}
H =  \sum_{<n,m>}
e^{iA_{\vec{n},\vec{m}}}c_{\vec{n}}^{\dag}c_{\vec{m}}\label{ham}
\end{equation}
\begin{equation}
\prod_{plaquette}e^{iA_{\vec{n},\vec{m}}}=e^{i\Phi}
\end{equation}
where $\Phi=2\pi{P\over Q}$ is a flux per plaquette, $P$ and $Q$
are
mutualy prime integers. In the most conventional Landau gauge
$A_{x}=A_{\vec{n},\vec{n}+\vec{1}_x}=0,
A_y=\Phi n_{x}$ the  Bloch wave function is
\begin{equation}
\psi(\vec n)=e^ {i\vec k \vec n}\psi_{n_x}(\vec
k),\,\,\psi_{n}=\psi_{n+Q}
\label{psi}
\end{equation}
where $n_{x}\equiv n=1...Q$ is a coordinate of the magnetic cell.
With
these substitution the Schrodinger equation  turns into a  famous
one-dimensional  quasiperiodic difference equation  ("Harper's"
equation):
\begin{eqnarray}
\label{harper}
 e^{ik_x}\psi_{n+1}+e^{-ik_x}\psi_{n-1}
+2\cos(k_{y}+n\Phi)\psi_n=E\psi_n
\end{eqnarray}
The spectrum of this equation has $ Q$ bands and feels the difference
between rational and irrational numbers - if the flux is
irrational, the  spectrum is singular continuum - uncountable but
measure
zero set of points (Cantor set). If the flux is rational, then the
spectrum has
$Q$ bands.

To ease the reference we state the main result of the paper:

It is known that due to gauge invariance, the energy depends on a
single parameter $\lambda= \cos (Qk_x)+\cos (Qk_y)$. We find that the
spectrum at $\lambda= 0 $ ( "mid" band spectrum ) is given by the sum
of
roots $z_l$\begin{equation}
E=iq^Q(q-q^{-1})\sum_{l=1}^{Q-1}z_l,\label{energy}
\end{equation}
of the {\it Bethe-Ansatz } equations for the Quantum Group
$U_q(sl_2)$
\begin{equation}
{{z_l^2+q}\over {qz_{l}^2 +1}}=
q^Q\prod_{m=1,m\ne l}^{Q-1} {{q z_l-z_m}\over {z_{l}-q z_m}},\,
\,l=1...Q-1.
\label{ba}
\end{equation}
 with
\begin{equation}
 q=e^{{i\over 2}\Phi},
\label{q}
\end{equation}
Another version of the Bethe-Ansatz equations is presented in the end
of the
paper.
Solution of the model with anisotropic hopping ( when the coefficient in front
of $\cos$ in Eq.(4) differs from 2 is also available. It will be published
elsewhere.
The Quantum Group symmetry is more transparent in another gauge
 $A_{x}=-{\Phi\over 2}( n_{x}+n_{y})$, $A_y={\Phi\over 2}(
n_{x}+n_{y}+1)$.
In this gauge a discrete coordinate of the Bloch function $ \psi(\vec
n)=\exp {(i\vec p \vec n)}\psi_{n}(\vec p)$, turns to $n=n_x+n_y $.

It is defined in two magnetic cells :
$n=1...2Q,\,p_{\pm}=(p_x{\pm}p_y)/2 \in [0,\Phi/2]$. An equivalent
form
of the Harper Eq. (\ref{harper}) for $\psi_{n}$ is

\begin{eqnarray}
\label{harper2}
&& 2e^{{i\over 4}\Phi + ip_{+}} \cos ({1\over 2}\Phi n+ {1\over
4}\Phi
-p_{-})\psi_{n+1} +
\nonumber\\
&&2e^{-{i\over 4}\Phi - ip_{+}} \cos ({1\over 2}\Phi n- {1\over
4}\Phi
-p_{-})\psi_{n-1}
=E\psi_{n}
\end{eqnarray}
In the new gauge the "midband " $\lambda=0$ corresponds to the point
$ \vec
p=({1\over 2}\pi,{1\over 2}\pi)$ \footnote{The doubling of period in comparison
with original Harper's
equation is artificial: using a simple transformation (multiplying
$\psi_{n}$ by $e^{-i\Phi n^{2}/4}$) one comes to an equation with
coefficients
of period Q. We are indebted to
Alexandre
Abanov for clarifying connection between two gauges}.
The advantage of this  gauge is that the wave function turns into the
polynomial
with   roots
$z_m$
\begin{equation}
\Psi(z)=\prod_{m=1}^{Q-1} (z -z_m)
\label{polynom}
\end{equation}
 at the points $z=q^{2l}$
\begin{equation}
\psi_{n}=\Psi(q^{n})\label{psi}
\end{equation}

\section{Group of Magnetic Translations}

The wave function of a particle in a magnetic field forms a
represantation of
the
{\it group of magnetic translations} \cite{Z}: let generators of
translations be

\begin{equation}
T_{ \vec \mu} (\vec i)=e^{iA_{\vec i,\vec  i+\vec \mu}}\mid \vec
i><\vec  i+\vec \mu\mid
\end{equation}
They form the algebra
\begin{eqnarray}
\label{tr}
&&T_{\vec{\mu}}=T_{-\vec{\mu}}^{-1},\,
T_{\vec{n}}
T_{\vec{m}}=q^{-\vec{n}\times \vec{m}}T_{\vec{n}+\vec{m}},
\nonumber\\
&&T_yT_x=q^2T_xT_y, \,T_yT_{-x}=q^{-2}T_{-x}T_y...
\end{eqnarray}
with $q$ given by the Eq.(7).
The Hamiltonian (\ref{ham}) can be expressed
\begin{equation}
H=T_x+T_{-x}+T_y+T_{-y}
\label{ham2}
\end{equation}
This group is also equivalent to the Heisenberg-Weyl group: $ [\hat
p,\hat
q]=
i\Phi,\,T_x=\exp {\hat q},\,T_y=\exp {\hat p}$.
\section {Quantum Group}

The algebra $U_q(sl_2)$ ( a q-deformation of the universalenveloping
of the $sl_2$)
is generated by the elements $A,B,C,D,$ with the commutation
relations
\cite{KR},\cite{sk},\cite{Dr},\cite{J},\cite{FRT}
\begin{eqnarray}
\label{ABCD}
&&AB=qBA,\,BD=qDB,
\nonumber\\
&&DC=qCD,\,CA=qAC,
\nonumber\\
&&AD=1,\,[B,C]={{A^2-D^2}\over{q-q^{-1}}}
\nonumber\\
\label{ABCD}
\end{eqnarray}
The center of this algebra is a $q-$analog of the Casimir operator
\begin{equation}
c=\left({{q^{-{1\over 2}}A-q^{1\over 2}D}\over {q-q^{-1}}}\right)^2
+BC
\label{casimir}
\end{equation}
In the classical limit $q\rightarrow {1+{i\over 2}\Phi}$, the quantum
group
turns to the $sl_2$ algebra:  ${(A-D)/(q-q^{-1})}\rightarrow S_3,\,
B\rightarrow S_{+},\,C\rightarrow S_{-},\,c\rightarrow {\vec S}^2
+1/4$.

The commutation relations (\ref{ABCD}) are simply another way to
write the
Yang-Baxter equation
\begin{equation}
R_{a_1a_2}^{b_1b_2}({u/v})L_{b_1 c_1}(u)L_{b_2 c_2}(v)
=L_{a_1b_1}(u)L_{a_2b_2}(v)R_{b_1b_2}^{c_1c_2}({u/v})
\end{equation}
where  generators $A,\,B,\,C,\,D$ are matrix elements of the $ L$-
operator
\begin{equation}
\label{lax}
L(u)=\left[
\matrix{{{uA-u^{-1}D}\over {q-q^{-1}}}&u^{-1}C\cr
         uB&{{uD-u^{-1}A}\over {q-q^{-1}}}\cr}
\right]
\label{L}
\end{equation}
Here $u$ is a spectral parameter and $R$- matrix is the $L$- operator
in
the spin 1/2-representation. It is given by the same matrix (\ref{L})
with
elements: $\, A=q^{{1\over 2}\sigma_3},\,D=q^{-{1\over
2}\sigma_3},\,B=\sigma_{+},\,C=\sigma_{-}$, where $\vec {\sigma}$ are
the
Pauli matrices.

Finite dimensional representations (except some representations of
dimension
$Q$ ) of the $U_q(sl_2)$ can be expressed in
the
weight basis, where $A$ and $D$ are diagonal matrices:
\,$\,A=diag\,(q^{j},...,
q^{-j})$. An integer or halfinteger $j$ is the spin of
the representation, and $2j+1$ is its dimension. The value of the
Casimir
operator (\ref{casimir})
in this representation is given by the $q$-
analog of $(j+1/2)^2$
\begin{equation}
c=\left({{q^{j+1/2}-q^{-j-1/2}}\over {q-q^{-1}}}\right)^2
=[j+1/2]^2_q
\label{casimir2}
\end{equation}

Representations can be realized by polynomials $\Psi(z)$ of the
degree
$2j$:
\begin{eqnarray}
\label{reprs}
&&A\Psi(z)=q^{-j}\Psi(qz),\,D\Psi(z)=q^{j}\Psi(q^{-1}z),
\nonumber\\
&&B\Psi(z)=z(q-q^{-1})^{-1}\left(q^{2j}\Psi(q^{-1}z)-q^{-2j}\Psi(qz)\right)
\nonumber\\
&&C\Psi(z)=-z^{-1}(q-q^{-1})^{-1}\left(\Psi(q^{-1}z)-\Psi(qz)\right)
\end{eqnarray}
This again the $q$-analog of the  representation of the $sl_2$
algebra by a
differential operator:
\begin{equation}
S_3=z{d \over dz}-j,\,S_{+}=z(2j-z{d \over dz}),\, S_{-}={d \over dz}
\label{sl2}
\end{equation}

\section{Magnetic Translations as a special representation of the
Quantum
Group}
 Dimension of our physical space of states is $2j+1=Q$. This is a
very special
dimension when $q^{2j+1}=\mp 1$ for $P$ - odd (even). The Casimir
operator
(\ref{casimir2}) in this case is
\begin{eqnarray}
\label{casimir3}
&&c=-4(q-q^{-1})^{-2},\, for \,P-odd
\nonumber\\
&&c=0,\,for \,P-even
\end{eqnarray}
In this special case  \cite{RA},\cite{Sk2} representation of the
quantum
group  can be naturally (but not unambiguously \cite{F}) expressed in
terms
of Magnetic Translations. Say one may choose
\begin{eqnarray}
\label{trqr}
&&T_{-x}+T_{-y}=\pm i(q-q^{-1})B,
\nonumber\\
&&T_{x}+T_{y}= i(q-q^{-1})C,
\nonumber\\
&&T_{-y}T_{x}=\pm q^{-1}A^2,\,T_{-x}T_{y}=\pm qD^2
\end{eqnarray}
where the upper sign corresponds to an odd $P$ and the lower to an
even
$P$ (representation of $U_q(sl_2)$ in terms of different but related Weyl
 basis can be found in Ref.\cite{RA},\cite{Sk2},\cite{F},\cite{bazhanov}).
 It is straightforward to check that this representation obeys
commutation
relations (\ref{tr}) and (\ref{ABCD}) and gives correct
values (\ref{casimir3}) of the Casimir operator (\ref{casimir}).

The Hamiltonian (\ref{ham},\ref{ham2}) now can be expressed in terms
of the
quantum group generators
\begin{equation}
H=i(q-q^{-1})(C\pm B)
\label{ham3}
\end{equation}
whereas the Schrodinger equation becomes a difference functional equation
\begin{equation}
i(z^{-1}+qz)\Psi(qz)-i(z^{-1}+q^{-1}z)\Psi(q^{-1}z)=E\Psi(z)
\label{eq2}
\end{equation}
The original Harper's discrete equation in the form (\ref{harper2} )
at the point $\vec p=(\pi /2,\pi /2)$ can be obtained from the
functional
equation (\ref{eq2} ), by setting $z=q^l$ and $\psi_l=\Psi(q^l)$. The
advantage to use
the extention of  $\psi_l$ to a complex plane $z$ is that the
representation theory of the quantum group garantees  that in a
proper
gauge the extended  wave
function would be  polynomial (9).

In addition to  representation  (\ref{reprs}) having the the highest and the
lowest  weight , in the special
dimension   $q^{2j+1}=\pm 1$ there is a parametric family of representations
having in general
no highest or lowest weights  \cite{Sk2},\cite{bazhanov}. The parameter
describes the anisotropy of the hopping amplitude in the
Hamiltonian (1) or the strength of the potential (i.e.the coefficient
in front of $cos$) in the Harper Equation (4). In this case the wave function
is   not polynomial in any gauge. Nevertheless the Bethe Ansatz
solution of the anisotropic problem is also possible but acqueres much heavier
mathematical techique. We postpone it for a more extended paper.

\section {functional Bethe-Ansatz}
Among various methods of the theory of quantum integrable systems the
{\it
functional Bethe-Ansatz} \cite{Sk3} seems to be the most direct way
to
diagonalize the Hamiltonian (\ref{ham3}).

We know that the solution of Eq.(\ref{eq2}) is polynomial

\begin{equation}
\Psi(z)=\prod_{m=1}^{Q-1} (z -z_m)
\label{polynom}
\end{equation}
 at the points $z=q^{2l}$
\begin{equation}
\psi_{n}=\Psi(q^{n})\label{psi}
\end{equation}
 Let us substitute
 it
in the Eq.(\ref{eq2}) and divide  both sides by $\Psi(z)$. We obtain
\begin{eqnarray}
\label{eq3}
&&i(z^{-1}+qz)\prod_{m=1,m\ne l}^{Q-1} {{qz -z_m}\over {z- z_m}}
\nonumber\\
&&-i(z^{-1}+q^{-1}z)\prod_{m=1,m\ne l}^{Q-1} {{q^{-1}z -z_m}\over
{z-z_m}}=E
\end{eqnarray}
The l.h.s. of this equation is a meromorphic function, whereas the
r.h.s.
is a constant. To make them equal we must null all residues of the
l.h.s..
They appear at $z=0$,\,at $z=\infty$\, and at $z=z_m$.
The residue at $z=0$ vanishes automatically.

The residue at $z=\infty$\ is $-iq^Q+iq^{-Q}$. Its null determines
the
deqree of the polynom.

Comparing  the coefficients of $z^{Q-1}$ in the both sides of
Eq.(24),we obtainthe energy given by Eq.(5).

Finally, annihilation of poles at $z=z_m$ gives the  Bethe-Ansatz
equations
(\ref{ba}) for roots of the polynomial (9). We write them here in a more
conventional form by setting $z_l=\exp {(2\varphi_l)}$
\begin{equation}
{{\cosh(2\varphi_l-i{\Phi\over 4})}\over
{\cosh(2\varphi_l+i{\Phi\over 4})}}=
\mp \prod_{m=1,m\ne l}^{Q-1}{{\sinh(\varphi_l-\varphi_m+i{\Phi\over
4})}\over
{\sinh(\varphi_l-\varphi_m-i{\Phi\over 4})}}
\label{ba3}
\end{equation}
Another form of the Bethe-Ansatz equations is given in the next
section.\section{ miscellaneous results}
{\bf 1.Another Form of the Bethe-Ansatz Equations}

As we already mentioned the representation of the quantum group by
magnetic
translations is not unique. This means that there is another gauge
where
the wave function is a polynom.
Consider the gauge $A_x=-A_y=-\Phi n_x$. Then instead of Harper's
equation(\ref{harper}) we obtain an equivalent equation for the gauge
transformed
wave function $\psi_n \rightarrow \phi_n=\exp ( i{{\Phi}\over
2}n(n-1))\psi_n$
(for an odd $Q$ it respects the periodic conditions (3)). At $\lambda
=0$, it has the form
\begin{eqnarray}
\label{harper4}
e^{-i\Phi n  }\phi_{n+1}+e^{i\Phi (n-1)}\phi_{n-1}
-2\cos(n\Phi)\phi_n=E\phi_n
\end{eqnarray}
This choice of the gauge corresponds to another representation of the
quantum group by magnetic translations. Say for an odd $P$ we have
\begin{eqnarray}
\label{trqr2}
&&T_{-x}+T_{-y}=-i(q-q^{-1})q^{-{1\over 2}}BD,
\nonumber\\
&&T_{x}+T_{y}=- i(q-q^{-1})q^{-{1\over 2}}CA,
\nonumber\\
&&T_{-y}T_{x}=q^{-1}A^2,\,T_{-x}T_{y}=qD^2
\end{eqnarray}
Then the Hamiltonian  (\ref{ham2}) turns into  quadratic form in the
$U_q(sl_2)$ generators
\begin{equation}
H=-i(q-q^{-1})q^{-{1\over 2}}(CA+BD)
\label{ham4}
\end{equation}
The representation (\ref{reprs})  (for $ q^{2j+1}=-1$) now gives
another
functional equation\begin{equation}
z^{-1}\Psi(q^2 z)+q^{-2}z\Psi(q^{-2}z)-(z+z^{-1})\Psi(z)=E\Psi(z)
\label{eq4}
\end{equation}
which is identical to (\ref{harper4}) on the set of points
$z=q^{2l}$,
where $l=
0...Q-1$. The Bethe-Ansatz may be obtained in the similar way:
\begin{equation}
z^2_l=
q^Q\prod_{m=1,m\ne l}^{Q-1} {{q^2 z_l-z_m}\over {z_{l}-q^2 z_m}},\,\,
l=1...Q-1.
\label{ba4}
\end{equation}
The energy is given again by the sum of roots
\begin{equation}
E=-q(q-q^{-1})\sum_{l=1}^{Q-1}z_l,\label{energy4}
\end{equation}
Inspite of the difference Eqs.(\ref {ba4},\ref {energy4}) must be
equivalent to the  Eqs.(\ref {ba},\ref {energy}).

{\bf 2.Quadratic Form of Quantum Group Generators}

A limited number of other interesting  solvable discrete equations
may be
obtained
from a general quadratic form of quantum group generators.
Their"classical"  version ($q\rightarrow 1$) would be  differential
equations
generated by quadratic forms of $sl_2$ -generators (\ref{sl2}) These
differential equations are known in the literature as so-called
"quasi
exactly soluable " problems of quantum mechanics \cite{Turb},\cite
{Ush}.
A quadratic form can be considerd as trace of a monodromy matrix of
an
integrable model with nonperiodic boundary conditions
\begin{equation}
\label{tau}
\tau= tr K_{+}(u)L(su)K_{-}(u)\sigma_2L^t(su^{-1})\sigma_2
\end{equation}
where $K_{\pm}$  are c-number matrices - solutions of "reflective "
Yang-Baxter equations  (RYB) \cite{Sk4},\cite{Cher}.  These matrices
describe all boundary conditions consistent with integrability There
is a
3-parametric family of boundary matrices  $K_{\pm}$ which generates a
general quadratic form of $A,\,B,\,C,\,D$.
Becides, there is a parameter $s$ in (\ref{tau}) which one can
introduce
in the $L$-operator (\ref{lax}) preserving integrability. For a
particular

choice of boundary $K$-matrices and parameter $s$ $\tau$ is
proportional
to the hamiltonian (\ref{ham4}). Therefore, one can say, that the
Hofstadter problem is equivalent to an
integrable magnet of spin $(Q-1)/2$ on one site with a proper
boundary
condition.

{\bf 3.q-Analog of Orthogonal Polynomials}

There is an intriguing connection between  a wave function
(\ref{polynom})
and q-generalization of orthogonal polynomials \cite{AW}. They
satisfy the difference equation (q-analog of differential
hypergeometrical
equation)
 \begin{eqnarray}
\label{aw}
A(z)P_n(q^2z)+A(z^{-1})P_n(q^{-2}z)-(A(z)+A(z^{-1}))P_n(z)
=(q^{-2n}-1)(1-abcdq^{2n-2})P_n(z)
\end{eqnarray}
where $A(z)=(1-az)(1-bz)(1-cz)(1-dz)/((1-z^2)(1-q^2z^2))$ and
$a,b,c,d$ are
parameters and $n$ is a degree of the polynomial. Choosing
$c=-d=q$,$a=-b=0$  we arrive at the equation for the q-Hermite
polynomials $H_n^{(q)}$\begin{equation}
H_n^{(q)}(q^2z)-{z}^2 H_n^{(q)}(q^{-2}z)=q^{-2n}(1-{z}^2
)H_n^{(q)}(z)
\end{equation}
(these are polynomials in $z+z^{-1}$ of degree $n$).
 For an odd $Q$ at  $n=(Q-1)/2$ this yields a zero energy solution to
eq.(\ref{eq4}) $\,\Psi^{(E=0)}(iz)=z^{(Q-1)/2}H_{(Q-1)/2}^{(q)}(z)$.

Another choice $c=-d=q$,  $a=-b=q$  and then  the replacement $q$ by
$q^{1/2}$
gives the q-Legendre equation
\begin{equation}
{{1-q{z}^2}\over {1-{z}^2}}P_n^{(q)}(qz)+{{q-{z}^2}\over {
1-{z}^2}}P_n^{(q)}(q^{-1}z)=(q^{-n}+q^{n+1})P_n^{(q)}(z)
\end{equation}
Then, comparing with the Eq.(\ref{eq2}) we conclude that the zero
mode
solution is given by the  q-Legendre polynomial  $\Psi^{(E=0)}
(iz)=z^{(Q-1)/2}P_{(Q-1)/2}^{(q)}(z)$.
\\
\\
The  almost immidiate and the most interesting task now is  to solve
the
Bethe-Ansatz equations in the limit $P, Q \rightarrow \infty$ when
the flux
$\Phi/2\pi$ is irrational. In all previous examples of integrable
systems it
was always possible to derive an integral equation for a distribution
function of roots $z_l$. We hope that it would be also possible for
this
problem and  after all allows  to obtain fractal properties  of the
spectrum analytically.\\
\\
\\
\noindent
{\bf ACKNOWLEDGEMENTS}\noindent

We would like to thank  P.G.O.Freund, A.Abanov, A.Gorsky, A.Kirillov,
E.Floratos, J.-L.Gervais and J.Schnittger for interesting
discussions. A.Z
is grateful to the Mathematical Disciplines Center of the  University
of
Chicago for the hospitality and support. P.W. acknowledges the
hospitality
of Weizmann Institute of Science and Laboratoire de Physique
Theorique de
l'Ecole Normale Sup\'{e}rieure where this work was completed. This
workwas supported in part by  NSF under the Research  Grant
27STC-9120000.
 
\end{document}